# On the behavior of the bounds of the holographic theory for massive and massless particle systems

Nicolo' Masi

INFN & Alma Mater Studiorum - Bologna University,

Via Irnerio 46, I-40126 Bologna, Italy

masi@bo.infn.it

**Abstract**

The aim of the present dissertation is to analyze the meaning of the entropy bounds of the holographic sector once tested for statistical ensembles of particles, in order to deeper investigate the nature of these constraints and their mutual links. From the *Universal Energy Bound* (UEB) simple time constraints can be argued, which are manifestations of the discrete nature of the space-time and of the presence of ultimate space-time scales. From the combined effort of the UEB and of the *Holographic bound* by 't Hooft and Susskind, an entropy density bound as a function of temperature is achieved.

The holographic theory includes two fundamental entropy bounds: the Universal Energy Bound by Bekenstein and the Holographic one, generalized by Bousso in his light-sheet reformulation [1]. Here we want to explore the behavior of the two bounds once applied to thermodynamic particle ensembles, whose physical quantities and properties are computed in the standard statistical mechanics framework. In fact, with this approach we try to explore the statistical mechanics meanings, moving from the evidence of an ultimate space-time scale which emerges in terms of $\pi T$ units. Even if the two bounds, in principle, deal with different physical quantities - energy and geometrical surface - for different systems, with a little bit of algebra their space-time constraints nature can be enlightened and put together to obtain something new, i. e. a hybrid entropy bound which contains both.

Let's start from the statistical mechanics calculations [2–4], which we are going to introduce into the entropy bounds to examine their behaviors: for massive particles the kinetic energy density is usually expressed in this way:

$$\rho_{massive} = \int_0^\infty \varepsilon N(\varepsilon) \, d\varepsilon = \frac{g \cdot M^{3/2}}{\sqrt{2}\pi^2} \int_0^\infty \frac{\varepsilon^{3/2}}{e^{\frac{\varepsilon-\mu}{T}} \pm 1} d\varepsilon = \qquad (1)$$

$$= \begin{cases} -\dfrac{g \cdot 3 M^{3/2} T^{5/2} \text{Li}_{\frac{5}{2}}(-e^{\frac{\mu}{T}})}{4\sqrt{2}\pi^{3/2}} \xrightarrow{\mu \to 0} \dfrac{g \cdot 3(2\sqrt{2}-1) M^{3/2} T^{5/2} \zeta\left(\frac{5}{2}\right)}{16\pi^{3/2}} \\ \dfrac{g \cdot 3 M^{3/2} T^{5/2} \text{Li}_{\frac{5}{2}}(e^{\frac{\mu}{T}})}{4\sqrt{2}\pi^{3/2}} \xrightarrow{\mu \to 0} \dfrac{g \cdot 3 M^{3/2} T^{5/2} \zeta\left(\frac{5}{2}\right)}{4\sqrt{2}\pi^{3/2}} \end{cases} = \begin{cases} b_1 (MT)^{3/2} \cdot T \\ b_2 (MT)^{3/2} \cdot T \end{cases}$$

where $N(\varepsilon)$ is the particle population as a function of energy, $T$ the temperature, $g$ is the spin-statistics factor, $\varepsilon$ is the kinetic energy of the particle and $M$ the total mass of the system, $\mu$ is the modified chemical potential $\mu \equiv \mu_0 - m$, which formally accounts for the rest mass energy, $\text{Li}_{\frac{5}{2}}$ is the polylogarithm and $\zeta\left(\frac{5}{2}\right)$ the Riemann Zeta. The upper case in curly brackets stands for fermions, the lower for bosons. The numerical coefficients are absorbed into the here-called $b_i$ coefficients. The $\mu \to 0$ approximation, for light particles far from degeneration and high enough temperature, leads to the final results.

For massless particles, i. e. ultrarelativistic fermions and photons, being $\nu$ the energy, the formulas are instead:

$$\rho_{urel} = \int_0^\infty \nu N(\nu)\, d\nu = \frac{g}{2\pi^2} \cdot \int_0^\infty \frac{\nu^3}{e^{\frac{\nu}{T}}+1}\, d\nu = \frac{g \cdot 7\pi^2 T^4}{240} = b_3 T^4 \tag{2}$$

$$\rho_{photon} = \int_0^\infty \nu N(\nu)\, d\nu = \frac{2}{8\pi^3} \cdot \int_0^\infty \frac{4\pi \cdot \nu^3}{e^{\frac{\nu}{T}}-1}\, d\nu = \frac{\pi^2 T^4}{15} = b_4 T^4$$

Here we have assumed that $\mu \equiv 0$, that is a solid condition for massless bosons and also for ultrarelativistic particles [2]; this is further supported by the fact that we want to use these thermodynamical relations for non-open systems.

For the coefficients the relation $b_1 < b_2 < b_3 < b_4$ stands. The entropy densities $s$ for these physical systems can be derived as follows, in natural units, using the energy densities so far computed:

$$s = \frac{\rho(1+w)}{T} = \frac{4}{3} b_{3,4} T^3, \qquad w = 1/3 \tag{3a}$$

$$s = \frac{\rho+p}{T} - \frac{\mu n}{T} = \frac{\rho(1+w) - \mu n}{T} \xrightarrow{w \to 2/3} \frac{5}{3} \cdot \frac{\mp g \cdot 3M^{3/2} T^{5/2} \text{Li}_{\frac{5}{2}}\left(\mp e^{\frac{\mu}{T}}\right)}{4\sqrt{2}\pi^{3/2} T} - \frac{\mu n}{T} \xrightarrow{\mu \to 0} \frac{5}{3} b_{1,2} (MT)^{3/2} \tag{3b}$$

where $w$ is the coefficient of the equation of state of the energy-matter gas and, as well-known in literature, it is worth $1/3$ for photons and low-pressure ultrarelativistic particles in (3a), and $2/3$ for massive pressure-carriers particles in (3b). The equation (3b) is the Gibbs-Duhem relation $Ts = \rho + p - \mu n$. The choice of $\mu \to 0$ condition, which leaves only the kinetic part of the energy amount of the system – neglecting the rest mass contribution included in the chemical $\frac{\mu n}{T}$ term – it is the common condition for fermions and bosons and it corresponds to the maximum value of the entropy density for fermions which have $\mu \geq 0$ and a minimum value for bosons ($\mu \leq 0$) and antifermions [2,5].

The consequent $s/\rho$ ratio for the massless case is always integrals independent and quite simple:

$$\frac{s}{\rho} = \frac{1+w}{T} = \frac{4}{3} \cdot \frac{1}{T} \tag{4a}$$

For the massive case, if we want to take into account the total energy-mass density $\bar{\rho} = \rho - \mu n$, the ratio can be written as follows

$$\frac{s}{\rho - \mu n} = \frac{1}{T}\left[1 + \frac{w}{1 - (\mu n)/\rho}\right] \xrightarrow{\frac{|\mu n|}{\rho} \ll 1} \frac{1+w}{T} \xrightarrow{w \to 2/3} \frac{5}{3} \cdot \frac{1}{T} \tag{4b}$$

The formula (4a) for the massless case is the ordinary property of a blackbody radiation confined to a cavity. Formula (4b) represents the equivalent formula for massive non-relativistic particles. The condition $\frac{|\mu n|}{\rho} \ll 1$ gives the asymptotic value $\frac{1+w}{T}$, for "enough kinetic" systems. In addition, it can be noted that, in order not to reach the massless factor 4/3 in (4a), associated for example to a photon gas, it follows that in (4b) the kinetic energy density must satisfy $\rho > \frac{|\mu n|}{3w-1}$. This implies that the rest mass density (and other eventual degrees of freedom) must be always less than the kinetic energy density. So, ultimately, the chemical term is conceptually quite irrelevant for our purpose and $\frac{5}{3} \cdot \frac{1}{T}$ represents a good point of reference for the current reasoning. These general results so far obtained are now applied to the entropy bounds.

First of all we invoke the **Bekenstein UEB bound** [6,7,8], expressed in terms of densities – entropy density and total energy-mass density $\bar{\rho}$, which coincides with $\rho$ for the massless case – and reported in ordinary units with the fundamental constants, which is

$$s \leq 2\pi R \bar{\rho} \left(\frac{k_B}{\hbar c}\right) \tag{5a}$$

It can be rewritten, from a space point of view, for the radius $R$ of the system using the asymptotic common result (4):

$$R \geq \frac{1}{2\pi} \cdot \frac{s}{\bar{\rho}} \left(\frac{\hbar c}{k_B}\right) \simeq \frac{1}{2\pi} \cdot \frac{1+w}{T} \left(\frac{\hbar c}{k_B}\right) \tag{5b}$$

It could be seen as a tight space cutoff as a function of temperature: the smaller the sphere is, the higher temperature the system has and vice versa.

It must be remembered that the universal energy bound is always relevant for weakly self-gravitating isolated systems with spherical symmetry (and for gravitating systems in general, once defined the physical quantities in a significant way), and for these systems it is a much stronger bound than the holographic one [6].

Now we can compute the minimum necessary time to transfer matter-energy (and information) from one through the sphere diameter of a spherical physical system, from the simple consideration that $\tau_{light} = \frac{2R}{c}$ and therefore the general relations $\tau \geq 2R$ and $\tau > 2R$ stand, respectively, for massless and massive particles. Replacing $R$ in (5b) with $\tau$, it is obtained:

$$\tau \geq \frac{s}{\bar{\rho}} \cdot \frac{1}{\pi} \left(\frac{\hbar}{k_B}\right) = \frac{1+w}{\pi T} \left(\frac{\hbar}{k_B}\right) \stackrel{N.U.}{\Longrightarrow} \begin{cases} \tau \gtrsim \frac{5}{3} \cdot \frac{1}{\pi T}, & M \neq 0 \\ \tau \geq \frac{4}{3} \cdot \frac{1}{\pi T}, & M = 0 \end{cases} \tag{6}$$

It emerges that the result (6) is independent from the temporal bound by Hod [9,10], and the Hod TTT bound value could be reached when the entropy density was equal to the energy density over $T$. The photon gas (and also the ultrarelativistic particles gas) presents a 4/3 factor which is close to one indeed, similar enough to the minimal quantum $(\pi T)^{-1}$ of the TTT bound, whereas the massive particles gas (the result also worth for classical Boltzmann particles) is characterized by a larger factor – almost equal to 5/3 and in general between 4/3 and 5/3 – and a consequent greater time to transmit information and energy from one extreme of the system to the other. For $w$ values different from 1/3 and 2/3 the relation (6) is no longer valid, because for $w \to 0$, the dust case, we approach the pure rest mass diluted particles case where $\mu \to 0$ is clearly wrong; for $w \to 1$, instead, the luminal $\tau \geq 2R$ condition could not stand anymore.

Once again the temporal inequality (6) could be read as a manifestation of the spatial-entropy bound $s \leq \pi\lambda(\rho + p)$ (see [11, 12, 13] for details), i. e. as a presence of an ultimate space-time scale: when the dimension $2R$ of the system becomes so small to be compared to the quantum Compton wavelength $\lambda$ of particle constituents at the assigned thermodynamical conditions. In fact, for the minimum scale $\lambda^*$ [11] we have $\lambda^* = \frac{s}{\rho} \cdot \frac{1}{\pi(1+w)}$ and, using formula (4), we obtain again $\lambda^* = 1/(\pi T)$. From the UEB it's not possible to achieve the exact TTT bound, even if from relations (6) the conceptual seed of a fundamental $\tau - T$ (time-temperature) constraint clearly emerges, which is very close to the TTT bound.

We can go on analyzing the **Holographic bound** by 't Hooft and Susskind [14,15] in terms of entropy-energy and space-time properties of the statistical mechanics particle ensemble. On one hand we have the Bekenstein space bound (5a) $R \geq \frac{1}{2\pi} \cdot \frac{s}{\rho}\left(\frac{\hbar c}{k_B}\right)$, and we want to deduce a comparable relation from the holographic entropy bound in order to do a close comparison and establish an explicit link between the two bounds. The Holographic bound $S \leq \frac{A}{4}\left(\frac{k_B}{l_P^2}\right)$ can be expressed in terms of entropy density. If we assume a spherical matter-energy distribution of surface $A = 4\pi R^2$ and volume $V = \int \sqrt{|g_{RR}|}\, 4\pi R^2 dR$ for a fixed observer in a generic curved space-time, and we compute an estimate of the entropy density inequality of a spherical slice of thickness $\Delta R \to R$ (to analyze how the bound behaves in first approximation), we get:

$$s \leq \frac{A}{4V}\left(\frac{k_B}{l_P^2}\right) \approx \frac{\Delta A}{4\Delta V}\left(\frac{k_B}{l_P^2}\right) = \frac{8\pi R \Delta R}{4 \cdot 4\pi R^2 \Delta R \sqrt{|g_{RR}|}}\left(\frac{k_B}{l_P^2}\right) = \frac{1}{2\sqrt{|g_{RR}|}} \cdot \frac{1}{R}\left(\frac{k_B}{l_P^2}\right) \tag{7a}$$

and

$$R \leq \frac{1}{2\sqrt{|g_{RR}|}} \cdot \frac{1}{s}\left(\frac{k_B}{l_P^2}\right) \tag{7b}$$

Introducing the ultrarelativistic entropy density - energy density relation (3a) in the radius inequality (7b), a fundamental formula, which is comparable to Bekenstein's one in (5b), it is finally obtained:

$$R \leq \frac{1}{2\sqrt{|g_{RR}|}} \cdot \frac{T}{\rho(1+w)}\left(\frac{k_B}{l_P^2}\right) \tag{8}$$

Combining the two inequalities (5a) and (8), it can be noted that the radius of the distribution must satisfy

$$\frac{1}{2\pi} \cdot \frac{s}{\rho}\left(\frac{\hbar c}{k_B}\right) \leq R \leq \frac{1}{2\sqrt{|g_{RR}|}} \cdot \frac{T}{\rho(1+w)}\left(\frac{k_B}{l_P^2}\right) \tag{9}$$

Relation (9) is worth for particle ensembles which satisfy the UEB energy bound and, secondarily, the Holographic entropy bound at the same time, i. e. for spherical self-gravitating non-open systems. In other words, the Bekenstein's bound represents a lower limit for the space-time $R - \tau$ scale, corresponding to the quantum Compton wavelength $\lambda$ of particle constituents, whereas the Holographic one is the upper limit beyond which the collapse occurs: they encode different information. In fact the UEB is a gravity-independent quantum bound, while the holographic bound has a meaning only in presence of gravity.
To reconcile the two side of the equation, the Bekenstein term on the left must be less or equal than the holographic one on the right. Thus we easily arrive at:

$$s \leq \frac{1}{\sqrt{|g_{RR}|}} \cdot \frac{\pi T}{1+w} \left(\frac{k_B}{l_P}\right)^2 \cdot \frac{1}{\hbar c} \stackrel{N.U.}{\Longrightarrow} \frac{1}{\sqrt{|g_{RR}|}} \cdot \frac{3}{4} \cdot \pi T \qquad (10)$$

This represents an *entropy-temperature bound* (ETB) for the holographic sector, which is a synthesis of the two holographic theory pillars. In addition, rewriting (10) in the following proper way

$$\frac{1}{s\sqrt{|g_{RR}|}} \geq \frac{4}{3} \cdot \frac{1}{\pi T} \qquad (11)$$

it can be put in evidence that, regardless the exact value of the coefficient (which depends on the definition of volume), the inverse of the entropy density is again proportional (and in general greater for flat space-time) to the $1/(\pi T)$ quantum.

The *entropy-temperature bound* carries the stringent nature of the UEB: in fact, if we compare it with the bound for a black hole [16], only for illustrative purpose, it becomes clear that it brings stronger conditions than the universal Holographic bound, which states:

$$S \leq A/4 = S_{BH} = \pi R^2 = 4\pi M^2 \qquad (12)$$

Rewriting (7a) for an euclidean volume $V = \frac{4}{3}\pi R^3$ and formula (10) for entropy $S = sV$, and not for entropy density, we get $S \leq \frac{2\pi^2}{1+w} R^3 T$. Then, replacing $R$ and $T$ with the Schwarzschild black hole relations $R = 2M$ and $T = 1/8\pi M$, where $R$ is the Schwarzschild radius, $T$ the BH temperature corresponding to the events horizon and $M$ the mass, the hybrid bound turns into

$$S \leq \frac{2\pi}{1+w} M^2 = \frac{3}{2}\pi M^2, \qquad w = 1/3 \qquad (13)$$

which is tighter than the pure holographic one in (12). The difference with respect to the holographic bound is $\frac{5}{2}\pi M^2$.

Another way to directly compare the mixed bound in (10) with a pure holographic one for black holes is the following. Hawking's black hole temperature $T$ can be expressed in terms of surface gravity $\kappa$ [17]: $T = \frac{\kappa}{2\pi}$. We know from Schwarzschild black hole theory that $\kappa = \frac{1}{4M} = \frac{1}{2R}$; using this relation in the previous, we obtain:

$$R = \frac{1}{4} \cdot \frac{1}{\pi T} \qquad (14)$$

This is once again a radius-temperature relation comparable with the inequality derived from the holographic bound. If we put (14) in (7b), another entropy density-temperature relation can be argued:

$$s \leq \frac{1}{\sqrt{|g_{RR}|}} \cdot 2\pi T \qquad (15)$$

This inequality is less stringent than the *ETB* in (10): we have double the $\pi T$ quantum, whereas the coefficient in (10) is less than one. So, this maximum entropy density, which finds correspondence into the universal holographic bound, can be reached, as expected, only by black holes systems and not by particles ones.

The presented results can be summarized as follows:

I.  The holographic sector main bounds can be analyzed in a statistical mechanics framework, producing simple space-time constraints such as (6) and (7b); the time-temperature bound (6) demonstrates the ubiquitous necessity of a minimum scale: it goes in the direction of the spatial-entropy bound by Pesci and, consequently, towards the TTT bound by Hod, even if it is not capable to saturate them. In fact, usual particle statistics cannot reach the extremal values [18].
II. The UEB and the Holographic bound are comparable, once put in evidence their complementary role, as shown in (9). They encode different extremes – microscopic and macroscopic respectively – of the space-time "ductility" towards the matter-energy content.
III. From the UEB and from the union of the two holographic bounds new inequalities for time and entropy density have been enlightened: they are manifestation of the $\pi T$ quantum, which underlies the space-time discretization.

I would like to thank Alessandro Pesci for comments and remarks on earlier versions of this manuscript.

## Bibliography

[1]  R. Bousso, arXiv:hep-th/0203101v2.
[2]  W. Greiner et al., *Thermodynamics And Statistical Mechanics*, Springer 1997.
[3]  D. A. McQuarrie, *Statistical Mechanics*, University Science Books, 2000.
[4]  K. Huang, *Statistical Mechanics*, John Wiley and Sons, 1987.
[5]  B. H. Lavenda, *A New Perspective of Thermodynamics*, Springer 2010.
[6]  J. D. Bekenstein, arXiv:quant-ph/0404042v1.
[7]  J. D. Bekenstein, Phys. Rev. D **7**, No. 8, 1973.
[8]  J. D. Bekenstein, Phys. Rev. D **23**, No. 2, 1981.
[9]  S. Hod, Phys. Rev. D **75** (2007) 064013.
[10] S. Hod, Phys. Rev. D **78**, 084035 (2008) , arXiv:0811.3806.
[11] A. Pesci, arXiv:1108.5066v1.
[12] A. Pesci, arXiv:0803.2642v2.
[13] A. Pesci, arXiv:0807.0300v2.
[14] G. 't Hooft, *Dimensional reduction in quantum gravity*, in Salam–festschrifft, A. Aly, J. Ellis, and S. Randjbar–Daemi, eds. (World Scientific, Singapore, 1993), arXiv gr–qc/9310026.
[15] L. Susskind, J. Math. Phys. **36**, 6377 (1995).
[16] T. Padmanabhan, *Gravitation – Foundations and Frontiers*, Cambridge University Press 2010.
[17] R. Wald, General Relativity, *University of Chicago Press*, 1984.
[18] N. Masi, arXiv:1109.4384.